# A Robust Hertz-Linewidth Quantum-Dot Coherent Swept Source with Adaptive Self-Linearization

Wei Chen[1+], Anyao Zhu[2+], Yueyang Zhang[1], Cunyu Shi[1], Tao Shu[1], Taijie Li[1], Gaopeng Wang[1], Zhile Wu[1], Liang Tang[1], Ying Yu[2*], Chenlei Li [1*], Daoxin Dai[1,3,4*]

*[1]State Key Laboratory for Extreme Photonics and Instrumentation, Zhejiang Key Laboratory of Optoelectronic Information Technology, College of Optical Science and Engineering, International Research Center for Advanced Photonics, Zhejiang University, Zijingang Campus, Hangzhou 310058, China;*

*[2]State Key Laboratory of Optoelectronic Materials and Technologies, School of Electronics and Information Technology, Sun Yat-Sen University, Guangzhou 510275, China;*

*[3]China Jiliang University, Hangzhou 310018, China;*

*[4]Intelligent Optics and Photonics Research Center, Jiaxing Research Institute, Zhejiang University, Jiaxing 314000, China.*

*+These authors contributed equally to this work.*

**Corresponding Author:* dxdai@zju.edu.cn; lichenlei@zju.edu.cn; yuying26@mail.sysu.edu.cn

**Abstract**

Frequency-modulated continuous-wave (FMCW) techniques underpin both coherent optical ranging and microwave radar, creating a common demand for highly coherent and highly linear frequency-chirped sources across the optical and microwave domains. Conventional semiconductor lasers, however, are fundamentally constrained by trade-offs among linewidth, linear tunability, and operational robustness. Here, we present a robust self-linearized quantum-dot (QD) coherent swept-source architecture through the co-design of source physics and system control. The chaos-free characteristics of QD lasers enable stable low-quality-factor external-cavity locking, yielding a Lorentzian linewidth of 12.6 Hz. Unlike self-injection-locked lasers, broadband external optical feedback allows the laser to maintain high optical coherence and stable operation over a wide current-tuning range, thereby enabling robust turnkey operation together with a chirp bandwidth of 23.2 GHz. Furthermore, a statistically gated real-time iterative learning control (ILC) strategy reduces the chirp nonlinearity ($1-R^2$) to as low as 8.96×10−8, while maintaining excellent environmental stability under laser-temperature variations. To demonstrate practicality, we demonstrate isolator-free coherent LiDAR and photonic generation of frequency-agile, linearly chirped microwave waveforms, establishing a common FMCW source platform for optical ranging and radar waveform synthesis. To the best of our knowledge, we demonstrate for the first time a QD-based coherent swept source that simultaneously combines hertz-level linewidth and ultrahigh chirp linearity, which we envision as a unified source platform for FMCW signal generation and coherent sensing across the optical and microwave domains.

## 1. Introduction

Frequency-modulated continuous-wave (FMCW) sensing encodes distance and velocity in the frequency difference between a transmitted chirp and its delayed return[1–3], linking measurement performance directly to the coherence and frequency trajectory of the source[4–6]. A narrow laser linewidth sustains coherence over long propagation delays[7–10], whereas a broadband, highly linear frequency sweep enables fine range resolution[11–13] and faithful recovery of target information. These properties must be maintained in the presence of optical feedback and thermal perturbations[4,14], which can destabilize laser operation and distort the sweep trajectory. For instance, temperature variations can shift the laser frequency[15,16] and external-cavity resonances[9], potentially disrupting locking in resonance-sensitive architectures. They can also alter the current-to-frequency response[17], causing previously calibrated driving waveforms to produce distorted chirps and thereby degrading ranging precision. The central challenge is therefore to combine long coherence with precisely controlled frequency agility in a source that remains robust under changing operating conditions. Meeting this challenge would advance coherent LiDAR and, through optical heterodyne conversion[18,19], extend the same source capabilities to linearly chirped microwave generation.

Recent advances in integrated semiconductor laser technologies have been substantial. Self-injection-locked (SIL) lasers can achieve an ultranarrow Lorentzian linewidth by locking the laser emission to a high-quality-factor (high-$Q$) microcavity. Previous studies have reported hertz-level Lorentzian linewidths on ultralow-loss silicon nitride ($Si_3N_4$) platforms[10,20–22]. Despite their remarkably low phase noise, SIL lasers are highly sensitive to operating conditions, as precise adjustment of the injection current is required to align the lasing frequency with a specific cavity resonance [14,21,23,24]. This not only increases operational complexity and susceptibility to environmental perturbations, but also constrains the achievable mode-hop-free tuning range. Alternatively, the emergence of thin-film lithium niobate (TFLN) has enabled ultrafast chirping via the intrinsic Pockels effect[25–28], reaching EHz $s^{-1}$ with chirp nonlinearity ($1-R^2$) below 1%[31]. However, the utility of such high-speed modulation in LiDAR is often constrained by the maximum unambiguous range (MUR)[4,29], because the required detection range imposes a lower bound on the chirp period. Furthermore, the direct-current drift of lithium niobate[30] often necessitates additional closed-loop control to maintain stable optical-frequency operation, while its material anisotropy can also constrain lateral chip scaling[31–33] and the realization of narrow-linewidth lasers. A swept-source architecture that combines narrow linewidth with broadband tuning, without relying on stringent resonance tracking, would therefore be valuable for robust FMCW sensing.

Direct current modulation of a semiconductor laser offers a simple and efficient approach to frequency sweeping[34]. However, their disparate response times can introduce dynamical hysteresis and distort the instantaneous frequency trajectory. Such chirp distortion broadens the heterodyne beat spectrum and degrades the fidelity with which range information can be recovered, becoming particularly detrimental as the propagation delay increases. To enable long-range detection, the $1-R^2$ of an FMCW LiDAR source typically needs to be suppressed to the level of $10^{-7}$ or below[13,35]. Previous studies[36,37] have shown that resampling can effectively correct chirp nonlinearity in post-processing. Although this approach avoids active linearization of the optical source, it requires additional computational processing and introduces latency. More fundamentally, non-uniform resampling modifies the original temporal coordinate of the detected waveform, which can distort or obscure Doppler information and therefore complicate simultaneous ranging and velocimetry[38]. Nevertheless, conventional implementations provide limited protection against noise accumulation, temperature drift, and time-dependent perturbations. Robust operation therefore requires a control strategy that can continuously distinguish systematic sweep distortion from stochastic measurement noise and adapt the driving waveform only when further correction is statistically justified.

Despite these advances, to our knowledge, no semiconductor FMCW source has yet demonstrated hertz-level linewidth, broadband frequency chirping with adaptive self-linearization, and robust operation under both strong optical feedback and temperature variations within a single architecture. Achieving this combination requires more than independent improvements in linewidth or chirp linearity:

the source must preserve stable coherent operation while its control scheme continuously compensates for changes in the frequency-sweep response. This calls for the co-design of an intrinsically resilient laser and an adaptive linearization strategy.

Here, we present a highly robust and self-linearized FMCW coherent swept-source architecture (Fig. 1a) based on a quantum-dot (QD) external-cavity laser (ECL), addressing the aforementioned limitations through the co-design of source physics and system control. The strong three-dimensional carrier confinement of the QD gain medium endows the laser with a linewidth-enhancement factor[23,24] that we previously measured to be as low as 0.136[39], stable lasing over a broad temperature range of 25–65 °C, and an optical-feedback tolerance of −10.2 dB. The chaos-free characteristics of QD lasers reduce the reliance on high-$Q$ microcavities for linewidth narrowing, enabling us to achieve a Lorentzian linewidth of 12.6 Hz using a filter-free low-quality-factor (low-$Q$) external cavity. Benefiting from broadband external optical feedback, the laser maintains high optical coherence and stable operation over a wide current-tuning range without stringent resonance alignment, thereby enabling robust turnkey locking together with a chirp bandwidth of 23.2 GHz. Furthermore, we develop a model-free real-time iterative learning control (ILC) strategy gated by statistical testing to adaptively compensate sweep distortions arising from carrier–thermal dynamics and environmental perturbations, reducing the $1-R^2$ to $8.96\times10^{-8}$ while maintaining long-term stable operation under laser-temperature variations. Enabled by this source, we further demonstrate isolator-free 100.1 km coherent FMCW fibre ranging, three-dimensional imaging, and Doppler velocimetry, with a single-point ranging precision of 711 μm. In addition, coherent heterodyne beating between two QD lasers enables widely tunable microwave-frequency synthesis and the generation of LCMW waveforms. These results provide a potential pathway towards future photonic LiDAR–radar integrated architectures for multi-domain sensing.

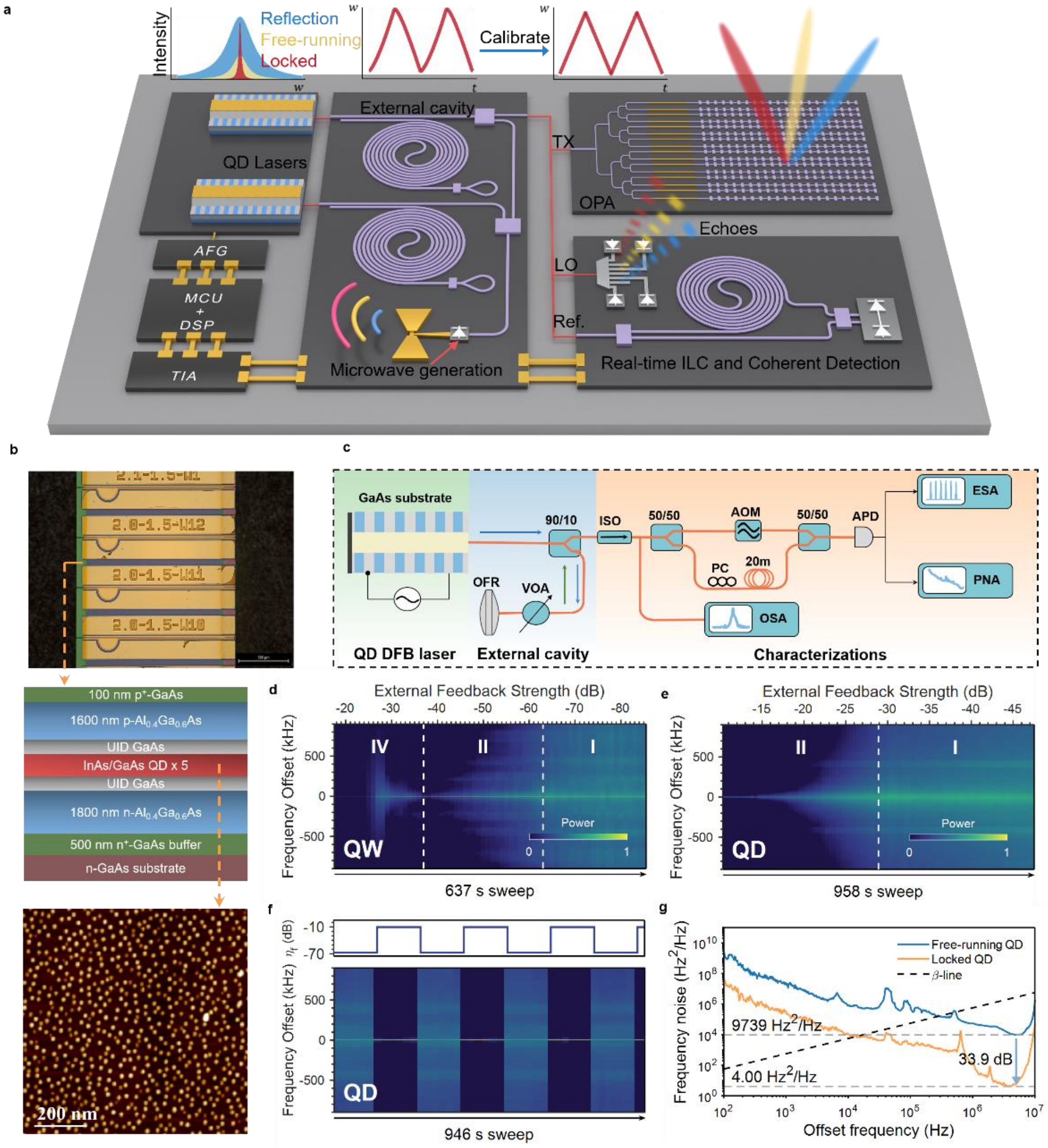

**Fig. 1 | QD coherent swept-source architecture and laser dynamics under external optical feedback.** **a,** Conceptual architecture of the proposed QD coherent swept-source platform for FMCW LiDAR and radar-waveform generation. Broadband low-$Q$ external-cavity feedback suppresses laser phase noise while preserving wide current-driven frequency agility, and real-time ILC calibrates the driving waveform to generate highly linear optical chirps. The calibrated swept source can be directed to an optical phased array (OPA) for FMCW LiDAR transmission and coherent detection, or heterodyned with a second QD laser to generate frequency-agile and linearly chirped microwave waveforms for radar. AFG, arbitrary function generator; TIA, transimpedance amplifier; MCU, microcontroller unit; DSP digital signal processor; TX, transmitter; LO, local oscillator. **b,** Microscope image and schematic epitaxial structure of the QD DFB laser array, together with a representative atomic force microscopy (AFM) image showing the uniform QD morphology. **c,** Experimental setup for external-cavity locking and linewidth characterization, including the off-chip external cavity and the self-heterodyne measurement system. OFR, optical fibre reflector; VOA, variable optical attenuator; ISO, optical isolator; AOM, acousto-optic modulator; PC, polarization controller; OSA, optical spectrum analyser; APD, avalanche photodiode; ESA, electrical spectrum analyser; PNA, phase-noise analyser. **d,e,** Experimentally measured maps of the spectral linewidth evolution of the commercial quantum-well (QW) laser and the

QD laser, respectively, as a function of external optical feedback strength, with the output power of both lasers fixed at 0 dBm. Regime I, stable operation regime; regime II, external-cavity-locking regime; regime IV, coherence-collapse regime. **f,** Turnkey operation of the QD laser under periodic modulation of the feedback strength between −68 dB and −10.2 dB. **g,** Frequency-noise spectra of the QD laser in the free-running (blue) and external-cavity-locked (orange) states, measured using the PNA.

## 2. Results

### 2.1 Robust QD ECL with turnkey locking

Robust coherent swept-source operation requires a gain medium with both high thermal stability and strong tolerance to external optical feedback. We therefore employ regrowth-free QD distributed-feedback (DFB) lasers directly grown on 3-inch GaAs (001) substrates as the foundation of the proposed source architecture (Fig. 1b). The discrete density of states in QDs suppresses carrier leakage[40] and reduces temperature sensitivity[41], enabling stable single-mode lasing over a broad operating-temperature range of 25–65 °C. At 65 °C and an injection current of 100 mA, the output power remains as high as 8.6 mW, corresponding to a reduction of only 0.98 dB from its peak value (Supplementary Fig. 1b). In addition, p-type modulation doping with Be in the barrier layers further suppresses thermally activated carrier escape, resulting in a nearly temperature-insensitive threshold current over 35–50 °C and a characteristic temperature $T_0$ as high as 149 K over 55–70 °C (Supplementary Fig. 1c). More detailed static laser characteristics and discussion of the thermal stability are provided in Supplementary Note I.

Coherent optical feedback in semiconductor lasers can increase the intracavity photon density and thereby substantially narrow the laser linewidth. Strong external optical feedback, however, may also induce severe coherence collapse[42]. Suppressing coherence collapse is therefore essential for fully exploiting the linewidth-narrowing benefit of external optical feedback. In addition to its high thermal stability, another key characteristic of the QD laser used here is its linewidth-enhancement factor (LEF) as low as 0.136, substantially lower than the typical value of 2–5 observed in quantum-well (QW) DFB lasers[43–45]. To evaluate how this intrinsic property translates into feedback-resilient coherent operation, we couple the QD laser to a low-$Q$ broadband external cavity and continuously vary the feedback strength using a variable optical attenuator (Fig. 1c). As a reference, a commercial QW laser with the same output power of 0 dBm is characterized under comparable feedback conditions (Fig. 1d).

Figure 1e summarizes the measured response of the QD laser as a function of feedback strength. When the feedback strength $f_{ext}$ is below −28.8 dB, the QD laser remains in the stable-operation regime (regime I), with its linewidth essentially unaffected by external feedback. As $f_{ext}$ increases from −28.8 to −10.2 dB, the laser enters the external-cavity-locking regime (regime II), accompanied by pronounced linewidth narrowing. In contrast, the commercial QW laser is substantially more sensitive to external optical feedback: it remains in regime II over a feedback range from −63.0 to −37.2 dB, whereas coherence collapse occurs once $f_{ext}$ reaches −37.0 dB. No coherence collapse is observed for the QD laser up to the strongest experimentally accessible on-chip feedback level of −10.2 dB, corresponding to an improvement in feedback tolerance of at least 26.8 dB relative to the QW laser. This enhanced feedback resilience primarily originates from the weaker amplitude–phase coupling and stronger damping associated with the low LEF of the QD gain medium[46]. A theoretical analysis of the relationship between the critical feedback strength, damping factor and LEF of semiconductor lasers under external optical feedback is provided in Supplementary Note II.

This strong feedback tolerance further enables robust turnkey external-cavity locking. When $f_{ext}$ is periodically modulated between −68 and −10.2 dB, the QD laser reproducibly switches between the free-running and locked states without active resonance tracking (Fig. 1f). The broadband external-cavity reflection relaxes the stringent resonance-alignment requirement associated with high-$Q$ resonant locking and enables stable coherent operation over a broad current-tuning range. At an injection current of 62 mA and a feedback strength of −10.2 dB, the white frequency-noise floor is reduced to 4.0 $Hz^2$ $Hz^{-1}$, corresponding to a 33.9-dB reduction relative to the free-running state and a Lorentzian linewidth of 12.6 Hz (Fig. 1g). These results demonstrate that combining the intrinsic feedback resilience of the QD laser with low-$Q$ broadband external-cavity locking simultaneously provides ultralow phase noise, strong feedback tolerance and stable coherent operation, thereby establishing the physical foundation for the self-linearized FMCW source developed below.

### 2.2 Real-time ILC algorithm for Temperature-robust FMCW linearization

The above material properties and laser-dynamics characteristics demonstrate that QD lasers are particularly suitable for achieving environmentally robust coherent operation. However, these intrinsic advantages must be further translated into a practical swept-source architecture. To improve the chirp linearity under realistic operating conditions, we develop a real-time ILC algorithm gated by statistical testing. The algorithm adaptively updates and optimizes the instantaneous frequency chirp based on the deviation between the measured and ideal frequency chirps, while incorporating the convergence behavior during the iterative process.

Figure 2a illustrates the workflow of the proposed algorithm. First, the instantaneous frequency chirp corresponding to the $k$-th iteration, $f_k(t)$, is extracted from the beat signal $U_k(t)$ through a Hilbert transform (HT):

$$f_k(t) = unwrap\left(\frac{arctan\left[\frac{HT\{U_k(t)\}}{U_k(t)}\right]}{2\pi\tau_{mzi}}\right) \quad (1)$$

Here, $unwrap(\cdot)$ is used to remove discontinuous phase jumps caused by $2\pi$ wrapping, and $\tau_{mzi}$ denotes the optical delay provided by the imbalanced Mach–Zehnder interferometer (iMZI), which is set to 2 m in this work, corresponding to a delay of 9.8 ns.

Second, we separate $f_k(t)$ into the up-ramp and down-ramp segments $f_{s,k}(t)$ andthe central 90% of each sweep is defined as the region of interest (ROI), within which a least-squares linear fit is performed to obtain the corresponding ideal linear frequency chirp $f'_{s,k}(t)$. The nonlinearity of the chirp is quantified by the goodness of fit $1 - R^2$, given by:

$$1 - R^2 = 12\left(\frac{e_{rms}}{B}\right)^2 \quad (2)$$

Here, $R^2$ is the coefficient of determination obtained from the least-squares linear fit within the ROI, $B$ denotes the measured chirp bandwidth, $e_{rms}$ denotes the root-mean-square (RMS) value of the residual between the frequency sweep trajectory and the reference straight line.The current waveform $u_{s,k}(t)$ is then updated according to:

$$u_{s,k}(t) = u_{s,k-1}(t) + \eta_{s,k} \cdot e_{s,k}(t) \quad (3)$$

Here, $s \in \{\text{up, down}\}$ denotes the up- and down-ramp segments, the tracking error between $f_{s,k}(t)$ and $f'_{s,k}(t)$ is normalized to the measured chirp bandwidth and mapped onto the corresponding driving-current range to obtain $e_{s,k}(t)$. The iterative learning gain $\eta_{s,k}$ is adaptively adjusted according to the achieved chirp linearity, while its initial value plays a critical role in determining the convergence rate of the algorithm, as discussed in Ref. [38]. In the steady-state and dynamic tests presented below, the initial iterative learning gain for both the up-ramp and down-ramp were set to 0.5, with a minimum threshold of 0.05. This setting enables rapid suppression of frequency sweep nonlinearity during the initial calibration stage while preserving good convergence behavior during steady-state operation.

It should be noted that, owing to the combined limitations imposed by the system transfer function and the stochastic nature of white noise[47,48], the true tracking error ultimately converges to a lower bound determined by the noise variance. Once the error approaches this limit, blindly continuing the iteration no longer improves the linearity[17], but instead may accumulate noise in the sweep waveform, thereby degrading the linearity and even causing algorithm divergence. To distinguish the origin of the error and avoid such ineffective convergence, we introduce a statistical decision mechanism into the real-time ILC. When $1 - R^2$ increases, the Pearson correlation coefficient between two successive error sequences, $e_{s,k-1}(t)$ and $e_{s,k}(t)$, is calculated and subjected to a significance test. In this work, the significance level $\alpha_{test}$ is set to 0.05. If the resulting p-value is lower than the prescribed $\alpha_{test}$, the two error sequences are considered significantly correlated, indicating that the observed degradation in linearity is mainly governed by deterministic factors, such as incompletely compensated thermal dynamics of the system. In this case, the current iteration is accepted and the current waveform is updated normally. Otherwise, the error sequences are regarded as uncorrelated, implying that the degradation is mainly caused by non-repetitive broadband background noise or transient environmental perturbations, and the current-waveform update is therefore rejected. A detailed analysis of the ILC algorithm is provided in Supplementary Note III.

To further characterize the modulation response, we generated a heterodyne beat signal between the locked DFB laser and a reference laser using a high-speed photodetector (Fig. 2a, navy). To maintain high output power and stable single-mode operation during current tuning, the 10% coupler port was used for feedback, corresponding to an on-chip feedback strength of approximately −19.7 dB. Through a bias tee, a triangular-wave drive with modulation frequencies ranging from 1 kHz to 100 kHz and an injection-current range of 40–100 mA was applied to the laser electrode. The heterodyne beat signal acquired in real time by the oscilloscope was analyzed using a short-time Fourier transform (STFT), and the deviation of the measured frequency trajectory from an ideal triangular-wave fit was then extracted. Under a 1 kHz triangular current ramp, the QD ECL achieves a chirp range of 23.2 GHz, whereas the chirp range decreases to 11.5 GHz when the modulation frequency is increased to 100 kHz. This reduction in the total chirp range mainly arises from the progressive suppression of the relatively slow thermal-tuning contribution, which provides a large frequency excursion but becomes increasingly attenuated at higher modulation rates[34]. In addition, without calibration, the intrinsic hysteresis of the thermal-tuning process gives rise to larger frequency deviations when the tuning direction switches between heating and cooling. This behavior manifests as asymmetric up- and down-sweeps, enhanced nonlinearity near the turning points, and higher-order harmonic components associated with spectral-energy spreading (Fig. 2b).

We first calibrated the chirp nonlinearity of the QD ECL at room temperature (25 °C) over repetition rates from 1 kHz to 100 kHz (Fig. 2c). Taking a repetition rate of 10 kHz as an example, the initial driving-current waveform was a standard triangular waveform spanning 40–100 mA. After iterative optimization, the maximum frequency deviation of the up-sweep was reduced from 1161 MHz to 7.45 MHz, while the $1-R^2$ within the ROI decreased from 0.021 to $1.05\times10^{-7}$. For the down-sweep, the maximum frequency deviation was reduced from 698 MHz to 6.61 MHz, with the corresponding $1-R^2$ decreasing from 0.019 to $8.96\times10^{-8}$.

To evaluate the robustness of the algorithm against environmental perturbations, we used an external thermoelectric cooler (TEC) to examine the long-term stability of the QD ECL under abrupt stage temperature variations. Over a total operating duration of 10.9 h, the temperature profile was divided into five stages (Fig. 2d): (1) stabilization at 25 °C for approximately 0.8 h; (2) heating to 50 °C over approximately 1.7 h; (3) stabilization at 50 °C for approximately 1.0 h; (4) cooling to 10 °C over approximately 4.6 h; and (5) returning to 25 °C over approximately 2.8 h. Figure 2e shows the evolution of $1-R^2$ and its standard deviation throughout the measurement. For clarity, each data point represents the mean $1-R^2$ within a 10 min time window. Under steady-state conditions, the mean values $1-R^2$ for the up- and down-sweeps are $1.74\times10^{-7}$ and $1.14\times10^{-7}$, respectively, after excluding data acquired during the first 10 min that did not pass the significance test. The enhanced perturbations during temperature transitions induce slight fluctuations in the operating linewidth of the QD external-cavity laser, resulting in a modest increase in the average $1-R^2$ compared with steady-state operation; nevertheless, it remains below $2\times10^{-7}$ throughout the measurement. After the significance test is first satisfied, the iterative learning gain remains close to its minimum threshold for most of the experiment, except for slight fluctuations during the heating processes in stages II and V (Fig. 2f).

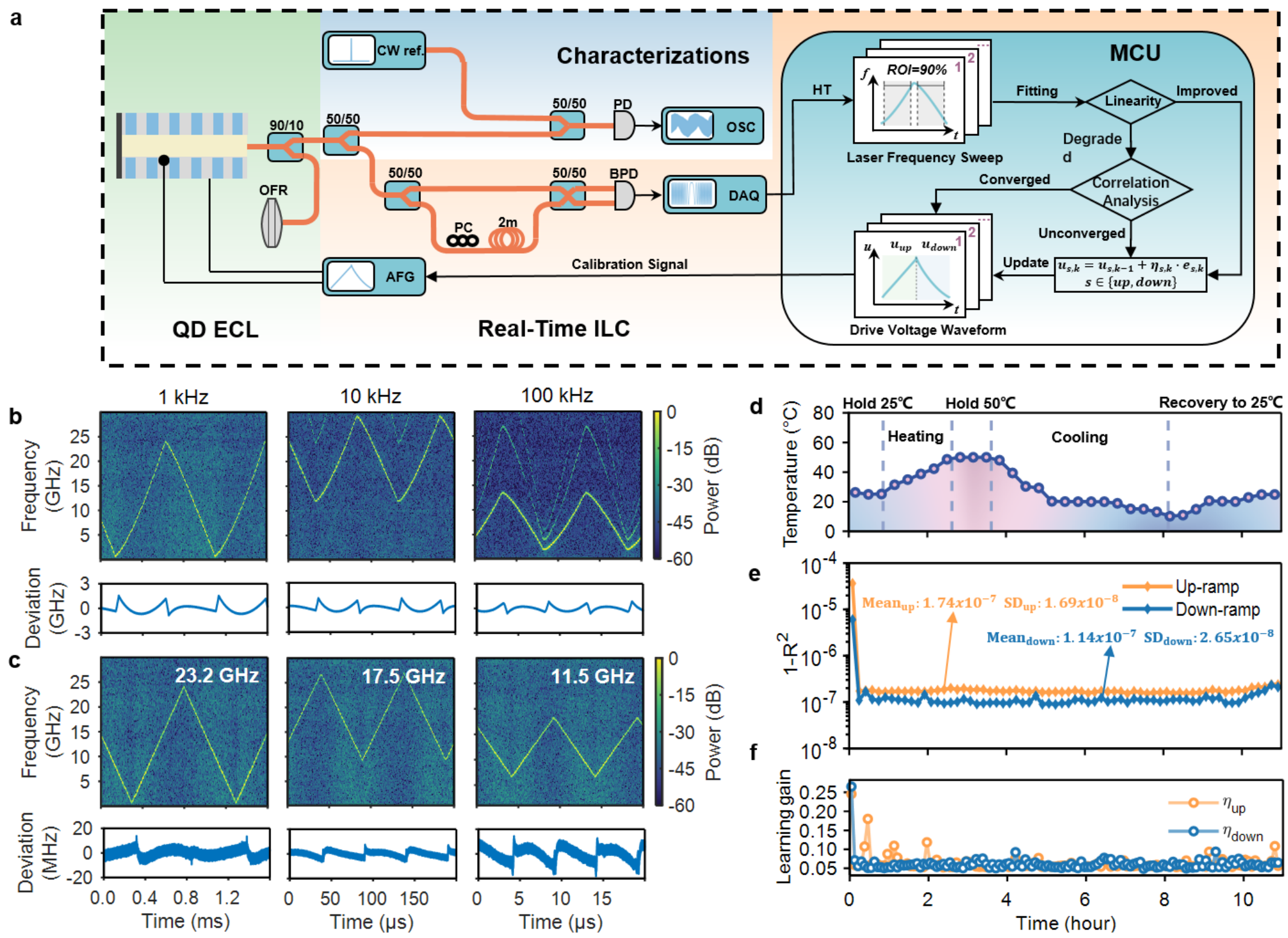

**Fig. 2 | Temperature-robust FMCW linearization and long-term stability. a,** Experimental setup for real-time ILC calibration and heterodyne beat-note measurement. CW ref., continuous-wave reference laser; PD, photodetector; BPD, balanced photodetector; OSC, digital oscilloscope; DAQ, data acquisition card. **b,c,** Upper panels: time–frequency spectra of heterodyne beat signals at modulation frequencies ranging from 1 kHz to 100 kHz before and after real-time ILC calibration. Lower panels: deviation between the measured frequency-sweep data and the least-squares linear fitting under the corresponding modulation frequencies before and after ILC calibration. **d,** Temperature evolution profile during long-term stability measurement using an external temperature controller, including five stages: stabilization at 25 °C, heating to 50 °C, holding at 50 °C, cooling to 10 °C, and returning to 25 °C. **e,** Evolution of the $1-R^2$ for the up-ramp and down-ramp processes during 10.9 h continuous operation. **f,** Dynamic evolution of the iterative learning gain $\eta_{s,k}$.

### 2.3 Versatile demonstrations enabled by the coherent swept source

#### 2.3.1 Coherent FMCW ranging

FMCW LiDAR retrieves range and velocity information by detecting the beat frequency between the reflected and reference optical fields, as illustrated in Fig. 4a. Based on the QD swept source developed above, we constructed the FMCW ranging system shown in Fig. 3a. The QD external-cavity laser was driven by a triangular waveform with a repetition rate of 1 kHz, corresponding to a chirp bandwidth of approximately 23.2 GHz. The system was evaluated under varying-temperature conditions (10–50 °C, Fig. 3b), where seven fibre lengths were measured 100 times each. The ranging precision, defined as the standard deviation of repeated measurements, is summarized in Fig. 3c. Figure 3d and e present the single-shot range resolution, defined as the full width at half maximum (FWHM) of the fast Fourier transform (FFT) spectrum peak, together with the corresponding distance-domain spectra. A range resolution of 15 cm is achieved at a fibre length of 2.1 km (inset of Fig. 3e).

For an ideal laser source, the range resolution of FMCW LiDAR is fundamentally determined by the speed of light $c$ and the total chirp bandwidth $B$, following $\delta R=c/2B$. However, in practical systems, the achievable resolution is additionally limited by two key properties of the laser source: phase noise and chirp linearity (Supplementary Note IV). Furthermore, as the measurement distance increases, the propagation delay becomes comparable to the duration of a single frequency sweep, reducing the effective temporal overlap between the transmitted chirp and delayed echo within the same sweep ramp. Consequently, the effective chirp bandwidth available for ranging is reduced.

Previous FMCW ranging demonstrations based on integrated sources have generally been limited to several tens of kilometres due to constraints in source coherence and sweep linearity. Benefiting from the ultranarrow linewidth and high chirp linearity of our swept source, the proposed system maintains high coherence beyond 100 km and achieves a signal-to-noise ratio exceeding 18 dB and a ranging precision of 7.091 m at a distance approaching the MUR (~102.13 km). At 100.1 km, the effective chirp bandwidth is reduced to only 0.5 GHz. As a result, the acquired beat signal inevitably contains highly nonlinear regions near the triangular-wave turning points, leading to a measured resolution that deviates substantially from the theoretical value (Fig. 3d). This limitation can be mitigated by employing oscilloscopes with higher sampling rates and larger bandwidths. Figure 3f compares the ranging performance of this work with previously reported long-range FMCW ranging demonstrations. The comparison highlights the ability of the present source to extend coherent ranging beyond 100 km while maintaining a competitive precision–distance trade-off, enabled by its narrow linewidth and real-time chirp linearization. It should be noted that the demonstrated 100.1 km range is not limited by source coherence: the 12.6 Hz Lorentzian linewidth corresponds to a theoretical $1/e$ fibre coherence length of 5150 km. The present range is instead constrained by the 102.13 km MUR and the 22.7 GHz beat frequency at 100.1 km. Lower chirp repetition rates and larger acquisition bandwidths could further extend the range towards several hundred kilometres.

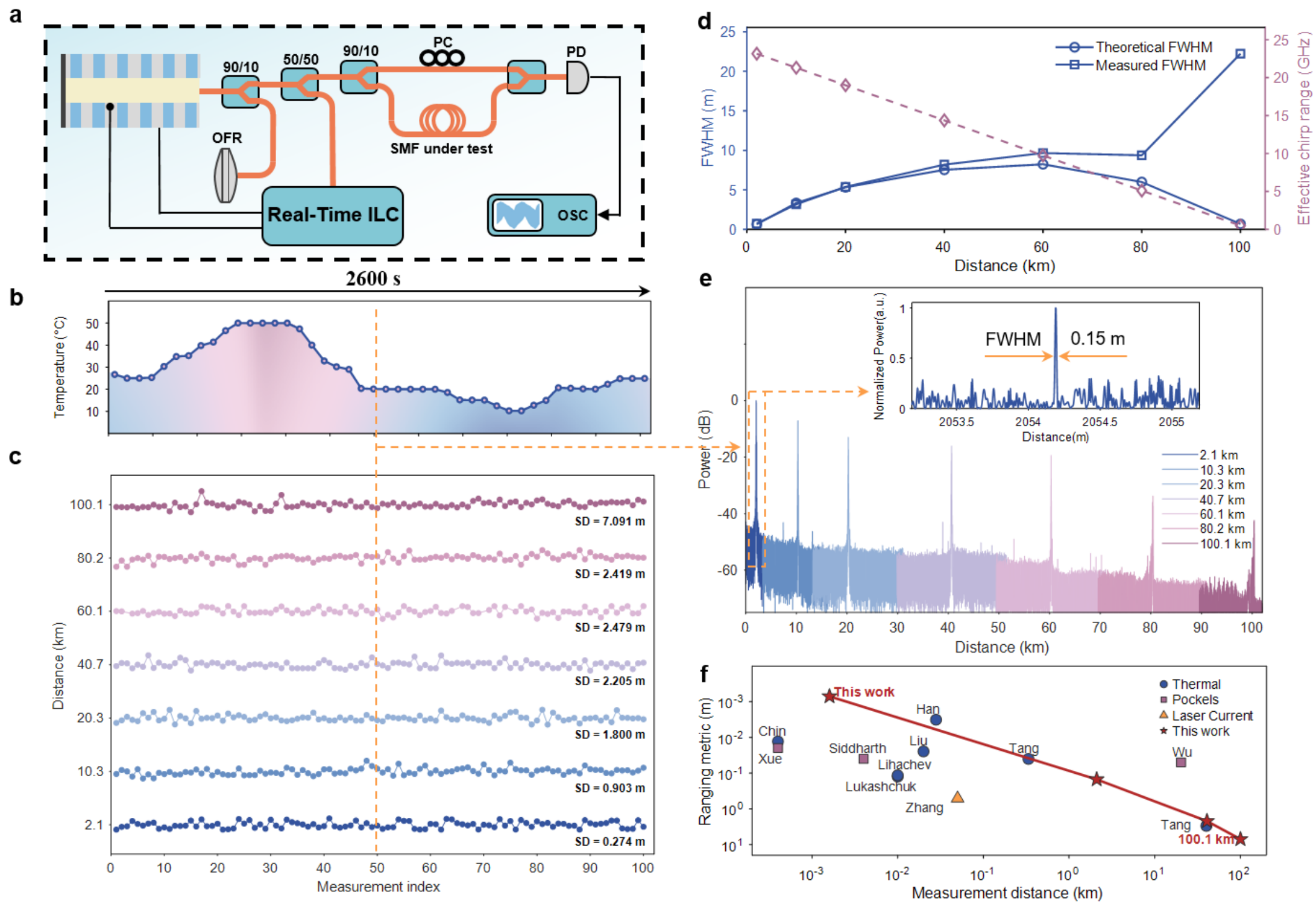

**Fig. 3 | Long-range coherent FMCW ranging. a,** Experimental setup for long-range FMCW ranging based on the QD swept source, in which part of the output is used for real-time ILC sweep linearization and the remainder is used for coherent ranging. **b,** Temperature variation profile during the measurement. **c,** Repeated ranging results for different fibre lengths under varying-temperature conditions, where SD denotes the standard deviation of 100 measurements. **d,** Theoretical and experimentally measured range resolutions, together with the effective chirp range, as a function of measurement distance. **e,** Representative distance-domain spectra corresponding to different fibre lengths from 2.1 to 100.1 km. **f,** Enlarged view of the main peak in the distance-domain spectrum for the 2.1 km measurement. **g,** Performance comparison between this work and previously reported long-range FMCW ranging systems[3,7,26,27,49–54].

### 2.3.2 Four-dimensional FMCW LiDAR sensing

To demonstrate the high-precision three-dimensional imaging and velocimetry capabilities of the proposed system, we constructed a coherent FMCW LiDAR setup (Fig. 4b). The QD external-cavity laser was driven by a triangular waveform with a repetition rate of 10 kHz, corresponding to a chirp bandwidth of approximately 17.5 GHz. A portion of the output light was used for real-time ILC-based sweep linearization, while the remaining power was directed to coherent detection. Notably, no additional optical isolator was inserted between the laser output and the booster optical amplifier (BOA), while the system maintained stable operation, further validating the isolator-free capability of the proposed source architecture. Both the transmitted and returned optical signals were coupled through the same collimator, which was combined with a circulator to provide directional isolation. The target scene was scanned using two galvanometers driven by triangular-wave signals. In the experiment, a hollow target plate and a diffuse-reflection background plate were selected as range-imaging targets. The former was positioned approximately 0.8 m away from the collimator, with a separation of approximately 0.8 m between the two targets. A flywheel with a diameter of 6 inches rotating at 600 rpm was employed to emulate a moving target, where different Doppler shifts were observed at different observation angles (photographs of the experimental targets are provided in Supplementary Note V).

The distance-domain spectrogram consists of 40,000 temporal slices. Figure 4c presents representative delayed beat signals between the local oscillator and the front and rear surfaces of the target at two different time frames, together with their corresponding signal-to-noise peak ratios (SNPRs) and full widths at half maximum (FWHMs). By extracting the center frequencies of the beat spectra, the corresponding range distributions were obtained and plotted as histograms in Fig. 4d. Double-Gaussian fitting further reveals the statistical characteristics of the two target-distance clusters. It should be noted that the standard deviations of the distance clusters in Fig. 4d are considerably larger than the theoretical range resolution, primarily due to the intrinsic curvature of the target surfaces and the buffer latency of the acquisition system.

To eliminate the influence of the foreground target, we removed the front panel and performed 1000 independent single-point measurements on the background panel located approximately 1.6 m from the collimator. A ranging standard deviation of 711 μm was achieved (Supplementary Note V). Based on the measured range information and the voltage–angle mapping relationship of the galvanometer controller, we reconstructed the three-dimensional optical ranging point cloud, as shown in Fig. 4e. The color of each point represents its distance relative to the collimator, clearly revealing the detailed features of the target pattern.

After validating static ranging and three-dimensional imaging, we further investigated the velocity measurement capability of the system. Figure 4f shows the two-dimensional velocity–frequency spectrum obtained at a modulation frequency of 100 kHz, where two distinct beat-frequency trajectories corresponding to the up-sweep and down-sweep processes ($f_u$ and $f_d$) can be clearly identified. These trajectories correspond to points on the flywheel with radial velocities ranging from −4 m s$^{-1}$ to 2 m s$^{-1}$. Figure 4g presents local time–frequency spectra extracted under representative target velocities, revealing clear velocity-dependent frequency shifts. The opposite motion directions result in frequency shifts with opposite signs, whereas the beat-frequency trajectories corresponding to the up- and down-sweeps nearly overlap when the target velocity approaches zero. By combining the velocity information extracted from the Doppler shift with the spatial ranging data according to the relation in Fig. 4a, we reconstructed a four-dimensional point cloud comprising three-dimensional spatial coordinates and radial velocity of the moving target (Fig. 4h). At a distance of 0.9 m from the collimator, the reconstructed point cloud not only accurately captures the spatial morphology of the target but also encodes its surface velocity distribution.

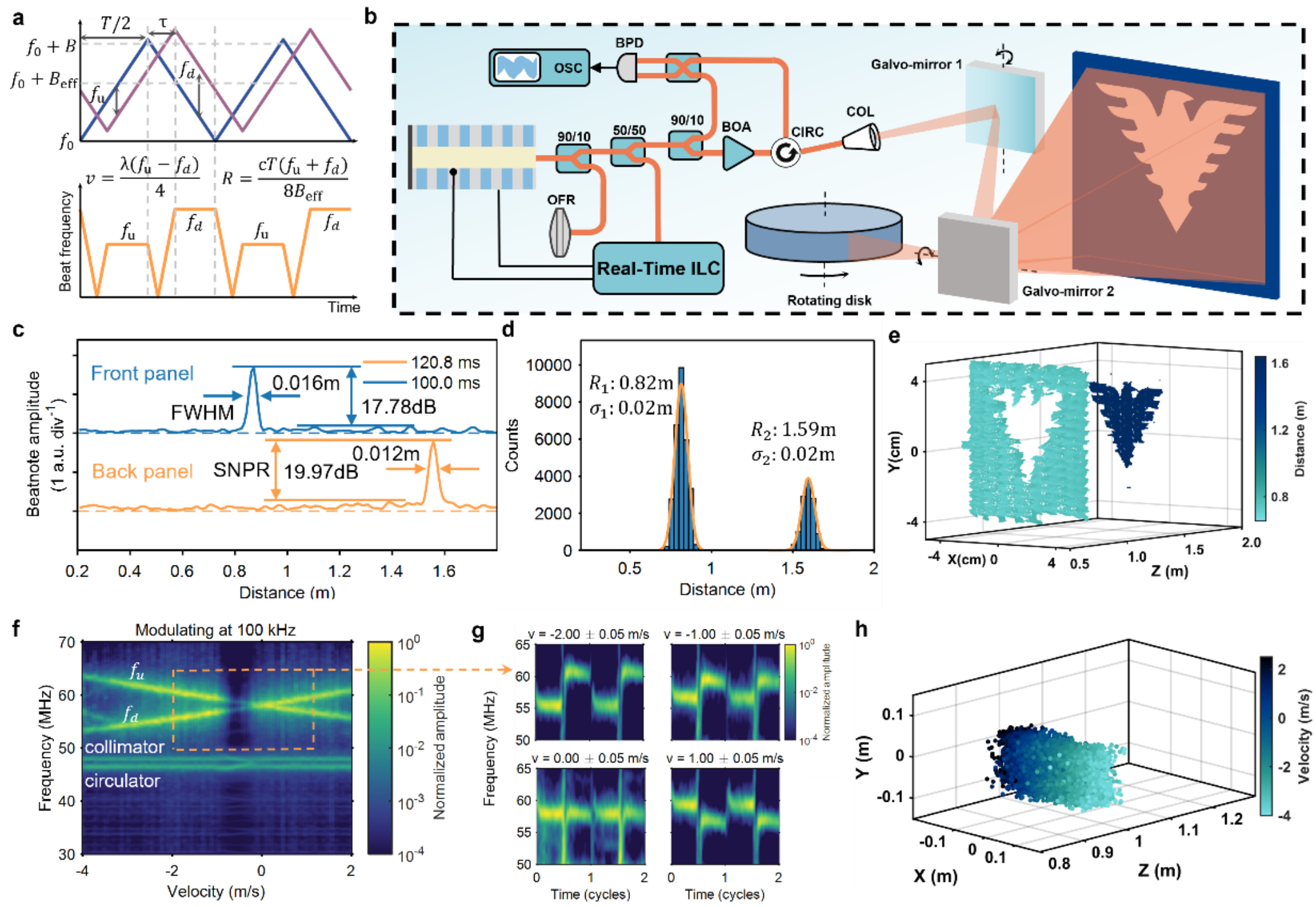

**Fig. 4 | Demonstration of coherent FMCW LiDAR. a,** Schematic illustration of the time-dependent optical frequency (upper panel) and beat frequency (lower panel) in FMCW LiDAR. **b,** Experimental setup for coherent FMCW LiDAR based on the QD swept source. BOA, booster optical amplifier; CIRC, circulator; COL, collimator. **c,** Representative delayed homodyne beat signal, with the corresponding SNPR and FWHM indicated. **d,** Histogram of the calculated target distance distribution, where both peaks are fitted using Double-Gaussian functions, with the extracted mean distance $d$ and standard deviation $\sigma$. **e,** Reconstructed three-dimensional point cloud of the target scene. **f,** Two-dimensional beat-frequency–radial-velocity spectrum of the rotating flywheel, where the fixed-frequency components originate from static reflections of the collimator and circulator. **g,** Representative time–frequency spectra of the beat signal at different radial velocities, with each dataset averaged over multiple FMCW cycles within the corresponding measurement range. **h,** Four-dimensional range–velocity point cloud of the rotating target, where the color represents the radial velocity extracted from the Doppler shift.

### 2.3.3 Tunable microwave generation

Photonics-based radio-frequency (RF) signal synthesis approaches can exploit the intrinsically broad optical bandwidth and operate directly at high frequencies[55,56]. The QD lasers demonstrated here integrate two sets of eight-channel local-area-network wavelength-division multiplexing (LAN-WDM) laser arrays on a single chip (Supplementary Note VI), enabling a wavelength separation exceeding 31 nm between two locked lasers, corresponding to a heterodyne frequency spacing beyond 5.6 THz. Therefore, direct heterodyne beating of the two optical sources on a high-speed photodetector provides a straightforward approach for frequency-agile microwave synthesis.

For direct on-chip microwave synthesis, feedback insensitivity is equally critical, as multiple components, including 3 dB couplers and photodetectors, need to be placed in close proximity to the lasers and can introduce strong on-chip optical reflections. To validate the feasibility of our lasers for heterodyne microwave synthesis, we performed the tunable microwave synthesis experiment shown in Fig. 5a. Long-term frequency stability can be further improved by implementing an optical phase-locked loop (OPLL) based on laser-current feedback (Fig. 5b), and this stability can be further enhanced through chip-scale packaging. The microwave frequency was tuned by varying the injection current of one QD ECL while maintaining a fixed injection current for the other laser. In such a heterodyne beating scheme, the phase noise of the generated microwave signal is determined by the combined phase noise of the two lasers involved in the beat process, while the achievable tuning range is ultimately limited by the bandwidth of the photodetector.

Figure 5c summarizes microwave signals generated from 0 to 44 GHz at 1-GHz intervals. The microwave frequency itself is continuously tunable. The tuning process is continuous, and the achievable frequency range can be further extended by employing photodetectors with larger bandwidths. The phase noise of the generated microwave signals at different frequencies was characterized, as shown in Fig. 5d. The results clearly indicate that the phase noise of the microwave signals is fundamentally determined by the phase noise of the two lasers involved in the heterodyne process, while the OPLL effectively suppresses low-frequency noise components.

Furthermore, direct heterodyne beating between one swept QD laser and another QD laser operated at a fixed injection current provides a straightforward approach for generating LCMW waveforms. Figure 5e presents the bandwidth of the generated LCMW signal. The acquired heterodyne beat signals were analyzed using a short-time Fourier transform, and the frequency trajectory was compared with an ideal 1 kHz triangular waveform. The extracted $1-R^2$ is as low as $3\times10^{-7}$ (Fig. 5f), corresponding to a time–bandwidth product of $1.134\times10^7$ for the generated LCMW waveform. Optical heterodyne microwave synthesis provides an effective approach for wide-range frequency-agile microwave generation at high frequencies, offering a practical pathway toward low-noise millimetre-wave and terahertz signal generation.

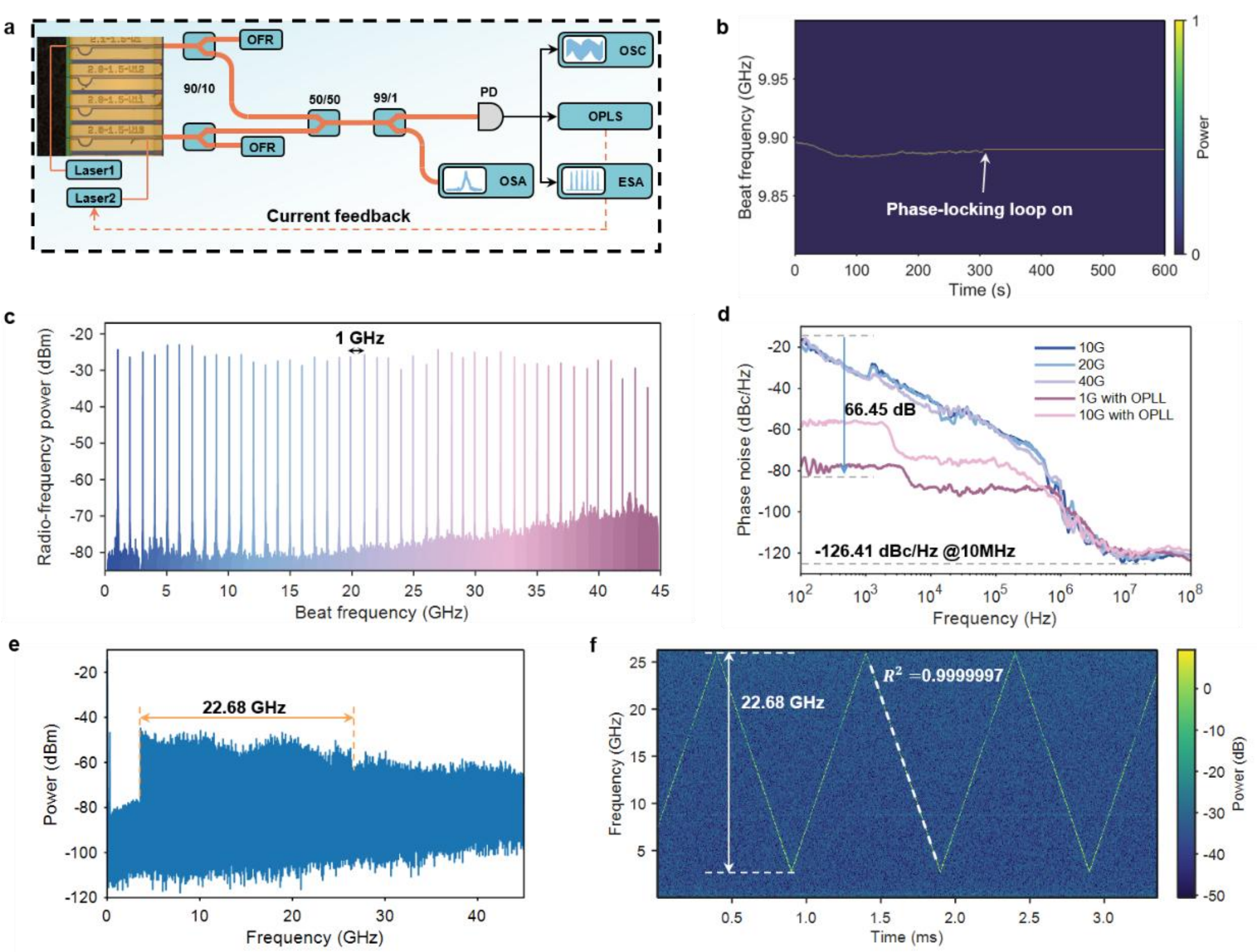

**Fig. 5 | Tunable microwave signal generation. a,** Experimental setup for isolator-free, widely tunable heterodyne microwave generation. OPLS, optical phase-locking servo. **b,** Long-term frequency stability

improvement of microwave generation enabled by the phase-locked loop. **c,** Generated microwave frequencies by tuning the wavelength of one QD laser while maintaining the other QD laser at a fixed wavelength. The maximum generated microwave frequency in this experiment is limited by the bandwidth of the PD. **d,** Carrier-frequency-independent phase noise of the generated microwave signals. The phase noise of the heterodyne microwave signal is directly determined by the two QD lasers involved in the beating process and is independent of the frequency separation between the lasers (i.e., the generated microwave frequency). **e,** Bandwidth of the LCMW signal generated by tuning one QD swept laser while maintaining the wavelength of the other QD laser fixed. **f,** Chirp nonlinearity of the generated LCMW signal.

## 3. Discussion

For FMCW sources intended for practical deployment, the key objective is not to maximize any single performance metric, but rather to achieve a synergistic balance among narrow linewidth, large chirp bandwidth, high chirp linearity, and environmental robustness. The regrowth-free QD laser employed in this work exhibits strong overall performance, including a threshold current of 19 mA, an SMSR of 56.5 dB, a fibre-coupled output power of 10.8 mW, and an output-power reduction of only 0.98 dB at 65 °C, demonstrating excellent temperature insensitivity. Its extremely low LEF substantially suppresses the conversion of intensity fluctuations into phase noise, allowing the laser to avoid coherence collapse even under optical feedback as strong as −10.2 dB, with the maximum feedback level limited by facet-coupling efficiency and insertion losses of the fibre components, and thereby enabling isolator-free operation. Compared with recently demonstrated Vernier-laser architectures requiring control of up to six analogue parameters, the low-$Q$ broadband external-cavity locking scheme requires only a single current control and avoids stringent narrowband resonance matching and precision tuning. Based on this architecture, the demonstrated QD external-cavity laser simultaneously achieves a Lorentzian linewidth of 12.6 Hz and a chirp bandwidth of 23.2 GHz, combining narrow linewidth, broadband chirping, and turnkey operation within a single source. Furthermore, the statistically gated real-time ILC developed in this work corrects the chirp nonlinearity without requiring an a priori model, reducing the nonlinearity to $8.96\times10^{-8}$ within 13 iterations under steady-state conditions, corresponding to an improvement by a factor of 212,053 over the uncorrected case. By combining the high-temperature single-mode stability and strong feedback resilience of the QD ECL with real-time ILC, the system maintains residual chirp nonlinearity at the $10^{-7}$ level under temperature variations and long-term operation. Without requiring an additional MZI for resampling or range correction, the system achieves sub-millimetre single-point ranging precision and further enables 100.1 km coherent fibre ranging, three-dimensional FMCW LiDAR imaging, Doppler velocimetry, and linearly chirped microwave waveform generation through heterodyne beating of QD lasers. In summary, we have proposed and experimentally validated a co-design paradigm for FMCW LiDAR sources targeting practical deployment, in which environmental robustness arises from the synergistic combination of intrinsic source-level resilience and robust calibration control. This approach provides a practical route towards translating coherent LiDAR from laboratory demonstrations to reliable operation in complex real-world environments.

Looking ahead, the proposed architecture could be further integrated on CMOS-compatible photonic platforms such as silicon and $Si_3N_4$, incorporating low-loss delay-line external cavities[21], on-chip calibration interferometers[17], coherent receiver chains[57], and control units within a unified platform. Such integration could substantially reduce the system footprint and insertion loss while improving packaging stability, thereby further alleviating the size, weight and power (SWaP) constraints of current

FMCW LiDAR systems. Meanwhile, the potential of these coherent swept sources, combining strong feedback resilience, wide-temperature stability, and real-time linearization, extends well beyond FMCW LiDAR. In industrial precision metrology, such sources could provide a promising platform for next-generation high-precision optical frequency-domain reflectometry (OFDR). In microwave photonics, their exceptional chirp linearity could support the generation of broadband, low-phase-noise LCMW and millimetre-wave signals. Furthermore, the combination of isolator-free operation, high-temperature tolerance, and monolithically integrated LAN-WDM multiwavelength capability could make this source architecture attractive for next-generation co-packaged optics (CPO), high-capacity coherent optical communications, and radio-over-fiber (RoF) networks, where robust on-chip light sources are increasingly required.

**Table 1. Performance comparison of representative frequency-swept laser sources.**

| Architecture | Ref. | Lorentzian linewidth(Hz) | Mode-hop-free Tunability (GHz) | Chirp nonlinearity | Feedback robustness (dB) | Temperature robustness (°C) | Ranging distance (m) |
|---|---|---|---|---|---|---|---|
| **QD-based Laser** | | | | | | | |
| DFB | 58 | $1.62 \times 10^3$ | n/a | n/a | > −14 | 25–85 | n/a |
| Low-$Q$ ECL | 24 | 16 | n/a | n/a | ≥ −9.6 | n/a | n/a |
| FP | 59 | n/a | n/a | n/a | −6.7 | 15–45 | n/a |
| Vernier | 60 | $50 \times 10^3$ | n/a | n/a | ≥ −10.6 | to 80 | n/a |
| **Low-$Q$ ECL** | **This work** | **12.6** | **23.2 GHz @ 1 kHz** | **$8.96 \times 10^{-8}$** | **≥ −10.2** | **10–50 FMCW** | **100,100** |
| **QW-based Laser** | | | | | | | |
| Ring SIL | 52 | 49.9 | 10.3 GHz @ 100 Hz | $3.0 \times 10^{-8}$ | n/a | n/a | 45,000 |
| Vernier | 49 | $2.9 \times 10^3$ | 7.68 GHz @ 1 kHz | $1.02 \times 10^{-7}$ | n/a | n/a | 20 |
| Ring SIL | 7 | 25 | >1 GHz | ~1%† | n/a | n/a | ~10 |
| Ring SIL | 28 | $3.14 \times 10^3$ | ~0.5 GHz linear chirp | ~1%† | n/a | n/a | 9.7 |
| E-DBR | 27 | 167 | 24 GHz @ 100 MHz | 3.8%† | n/a | n/a | 1 |
| E-DBR | 26 | $2.8 \times 10^3$ | 9.3 GHz @ 1 MHz | ~2.31%† | n/a | n/a | 4.5 |
| Vernier | 61 | n/a | 21.8 GHz @ 1 kHz | $6.01 \times 10^{-7}$ | n/a | n/a | 300 |
| Vernier | 51 | $1.2 \times 10^3$ | 62 GHz @ 50 Hz | $2.36 \times 10^{-9}$ | n/a | n/a | 300 |
| MPDF SIL | 25 | 10.6 | 3 GHz | $8.9 \times 10^{-6}$ | n/a | n/a | 3,000 |
| RE-DBR | 14 | 24 | 34 GHz (static) | n/a | n/a | 15–50 °C lasing | n/a |

† Relative RMS deviation. FP, Fabry–Pérot; E-DBR, extended distributed Bragg reflector; MPDF, multiperiod-delayed feedback; RE-DBR, resonator-enhanced distributed Bragg reflector; n/a, not available.

## 4. Method

### 4.1 Laser characterizations

The frequency-noise spectra and corresponding intrinsic linewidths of the laser in the free-running and locked states were characterized using a delayed self-heterodyne measurement. The laser was driven by a low-noise current source (Keithley 2230G-30-1), and its output was divided into two channels. The first channel was directed to an OSA (Yokogawa AQ6370C) to record the optical spectrum and verify single-mode operation. The second channel was sent to the self-heterodyne interferometer, where the generated beat signal was detected by an APD (E80040271) and subsequently analyzed using a PNA (Keysight E5052B). Fifty cross-correlation averages were performed over an offset-frequency range of 100 Hz–20 MHz to suppress the influence of the instrument noise floor. The delayed phase-noise spectrum $S_{\Delta\phi}(f)$ measured by the PNA was converted into the frequency-noise spectrum according to:

$$S_{\nu}(f) = \frac{f^2}{4sin^2(\pi f\tau)} S_{\Delta\phi}(f) \tag{4}$$

where $f$ is the offset frequency and $\tau$ is the relative delay between the two interferometer arms. The intrinsic linewidth was then extracted from the high-frequency white-noise floor $S_{v,0}$ using $\Delta\mathrm{v} = \pi S_{v,0}$. The self-heterodyne interferometer consisted of an unbalanced MZI, with a PC and 20 m delay fibre in one arm and an AOM (KG-AOM-1310-80M-SS-FA) driven at 80 MHz in the other arm.

Figure 1c shows the experimental setup for external-cavity locking. The coupled laser output was first sent to a 90/10 fibre splitter, with 90% of the coupled power used for external optical feedback. The feedback loop consisted of a PC, a VOA (Joinwit JW3301), and an OFR. The external-cavity frequency was 10.36 MHz, corresponding to a fibre length of 9.86 m. The remaining 10% of the laser output was used to characterize the feedback sensitivity and laser linewidth. All losses in the external cavity were included in the calculation of the reflected power. The fibre–chip coupling loss was −2.72 dB, corresponding to a round-trip loss of −5.44 dB, while the combined loss introduced by the fibre splitter, PC, VOA, OFR, and fibre connectors was −4.76 dB. Consequently, the maximum on-chip feedback strength achievable in the present setup was −10.2 dB. The locking state was identified by feeding the APD-detected beat signal into an ESA (Keysight N9020B), which was also used to characterize the linewidth-narrowing behavior under different feedback strengths (Fig. 1d, e).

The chirp range of the QD external-cavity laser was characterized using heterodyne beating with a wavelength-tunable narrow-linewidth benchtop laser (Santec TSL-550) as the reference source. Current modulation was applied to the laser under test through bias tees (Sensefuture VCS100 at 1 kHz and Thorlabs T1G at 10–100 kHz). The heterodyne beat signal between the two optical fields was detected using a high-speed photodiode (Albis PQS40A-L), and its electrical output was recorded by a real-time oscilloscope (Keysight DSA-X 93204A). The instantaneous dominant frequency was subsequently extracted using a short-time Fourier transform (Fig. 2b, c).

### 4.2 Four-dimensional FMCW LiDAR experiment and signal data processing

In the static three-dimensional imaging experiment, the laser was driven by a triangular waveform with a repetition rate of 10 kHz and a current range of 40–100 mA, corresponding to a chirp bandwidth of 17.5 GHz. For the coherent velocimetry experiment, the system configuration and signal acquisition procedure remained unchanged, except that the repetition rate of the triangular driving waveform was increased to 100 kHz while maintaining the same current range of 40–100 mA, corresponding to a chirp bandwidth of 11.5 GHz. The laser output power was amplified to approximately 8 dBm using a booster optical amplifier (Thorlabs S9FC1132P). The emitted beam was scanned across the target scene using two galvanometer scanners (Thorlabs GVS012), driven by triangular-wave signals generated by a DAQ (NI USB-6453) at 500 Hz in the vertical direction and 10 Hz in the horizontal direction. The horizontal and vertical scanning angles were obtained from the galvanometer driving signals and recorded by the same DAQ. To compensate for chirp nonlinearity, a portion of the output light was used for real-time ILC calibration. Since each iteration of the frequency-sweep waveform introduces a slight variation in the chirp bandwidth, the chirp bandwidth used for range calculation was dynamically updated after each iteration to eliminate ranging errors caused by bandwidth variation. The system employed a monostatic transceiver configuration, where the return signal was collected by a collimator with a beam waist diameter of 3.4 mm (Thorlabs F280APC-C) and mixed with the local signal to generate beat signals. The acquired data were processed using zero-padded STFT, followed by spectral filtering to remove frequency components corresponding to reflections from the fibre collimator, circulator, and fibre connectors. The fixed propagation distance between the laser and the collimator was subtracted, and the point-cloud data were calibrated with respect to the aperture position of the collimator.

### 4.3 Performance comparison of representative frequency-swept laser sources

We benchmark the quantum-dot coherent swept source proposed in this work against representative semiconductor laser sources reported in the literature. QD laser studies without an explicit frequency-sweeping demonstration are additionally included when they provide relevant benchmarks for linewidth, optical-feedback tolerance, or temperature robustness. The comparison metrics include the Lorentzian linewidth, experimentally demonstrated mode-hop-free tuning or chirp range, chirp nonlinearity, optical-feedback robustness, temperature robustness, and experimentally demonstrated ranging distance, as summarized in Table 1. The sources are categorized according to the gain medium (QD or QW) and laser architecture. For the tuning bandwidth, only experimentally demonstrated mode-hop-free tuning ranges or frequency ranges explicitly reported as continuous chirps are included; broad wavelength-tuning ranges achieved by switching longitudinal modes or changing operating points are excluded. Chirp-nonlinearity values are quoted directly according to the definitions adopted in the original publications, including $1-R^2$ and relative RMS deviation, and are therefore not renormalized across different reports. Optical-feedback robustness is defined as the strongest reported feedback level under which stable operation is maintained, or the experimentally measured coherence-collapse threshold, where available. Temperature robustness corresponds to the experimentally demonstrated temperature range for stable lasing or FMCW operation, as indicated in Table 1. The ranging distance refers to the longest distance experimentally demonstrated rather than a theoretically estimated maximum range. Parameters that were not reported or directly demonstrated in the original publications are marked as n/a.

**Funding.** This work was supported in part by the National Natural Science Foundation of China (62321166651,62405271,62550180), Zhejiang Provincial Natural Science Foundation of China (LD26F050001), Fundamental and Interdisciplinary Disciplines Breakthrough Plan of the Ministry of Education of China (JYB2025XDXM106), Fundamental Research Funds for the Central Universities (226202400171), China Postdoctoral Science Foundation( 2026M790888, 2026T190224)

**Disclosures.** The authors declare no conflicts of interest.

**Data availability.** Data underlying the results presented in this paper are not publicly available at this time but may be obtained from the authors upon reasonable request.